\documentclass[aps,prl,twocolumn,superscriptaddress,notitlepage,nofootinbib]{revtex4-1}
\usepackage{amsmath}
\usepackage{epsfig}
\usepackage{color,soul}
\usepackage{multirow}
\def\bea#1\eea{\begin{align}#1\end{align}}

\newcommand{\bef}{\begin{figure}[h!tb]\centering}
\newcommand{\eef}{\end{figure}}
\newcommand{\jsi}{J/\psi}

\begin{document}
\title{$\jsi$ production and polarization within a jet}

\author{Zhong-Bo Kang}
\affiliation{Department of Physics and Astronomy, University of California, Los Angeles, California 90095, USA}
\affiliation{Mani L. Bhaumik Institute for Theoretical Physics, University of California, Los Angeles, California 90095, USA}
\affiliation{Theoretical Division, Los Alamos National Laboratory, Los Alamos, New Mexico 87545, USA}
                   
\author{Jian-Wei Qiu}
\affiliation{Theory Center, Jefferson Lab, 12000 Jefferson Avenue, Newport News, Virginia 23606, USA}
                                
\author{Felix Ringer}
\affiliation{Nuclear Science Division, Lawrence Berkeley National Laboratory, Berkeley, California 94720, USA}
\affiliation{Theoretical Division, Los Alamos National Laboratory, Los Alamos, New Mexico 87545, USA} 

\author{Hongxi Xing}
\affiliation{Department of Physics and Astronomy, Northwestern University, Evanston, Illinois 60208, USA} 
\affiliation{High Energy Physics Division, Argonne National Laboratory, Argonne, Illinois 60439, USA}
                   
\author{Hong Zhang}
\affiliation{Department of Physics, The Ohio State University, Columbus, Ohio 43210, USA}

\date{\today}         

\begin{abstract}
We study the production and polarization of $\jsi$ mesons within a jet in proton-proton collisions at the LHC. We define the $\jsi$-jet fragmentation function as a ratio of differential jet cross sections with and without the reconstructed $\jsi$ in the jet. We demonstrate that this is a very useful observable to help explore the $\jsi$ production mechanism, and to differentiate between different NRQCD global fits based on inclusive $\jsi$ cross sections. Furthermore, we propose to measure the polarization of $\jsi$ mesons inside the jet, which can provide even more stringent constraints for the heavy quarkonium production mechanism. 
\end{abstract}

\date{\today}

\maketitle
{\it Introduction.} Understanding the $\jsi$ production mechanism is one of the most active and challenging subjects in strong interaction physics~\cite{Brambilla:2010cs,Bodwin:2013nua}. The most common approach to calculating the $\jsi$ production cross section in hadronic collisions is the non-relativistic QCD (NRQCD) factorization formalism~\cite{Bodwin:1994jh}. In this approach, the heavy quark and anti-quark pair, $Q\bar Q$, is produced at short-distance which can be calculated perturbatively due to the large heavy quark masses. Such a $Q\bar Q$ state will then hadronize into a physical $\jsi$ meson. This transition is non-perturbative, but it can be characterized through a set of universal NRQCD long-distance matrix elements (LDMEs). Global analyses of the world's data on $\jsi$ production with next-to-leading order (NLO) calculations in powers of strong coupling constant $\alpha_s$ have been performed by several groups~\cite{Butenschoen:2011yh,Chao:2012iv,Gong:2012ug,Bodwin:2014gia,Bodwin:2015iua}. Even though all these groups can describe the inclusive $\jsi$ production cross section, i.e., the $p_T$ spectrum, they have not been able to fully explain the polarization of high-$p_T$ heavy quarkonia 
produced at the Tevatron~\cite{Affolder:2000nn,Abulencia:2007us,Acosta:2001gv,Abazov:2008aa} and the LHC~\cite{Chatrchyan:2012woa,Chatrchyan:2013cla}. Despite numerous attempts made in the past, the $\jsi$ polarization remains a puzzle. 

In this Letter we explore novel opportunities to further study the $\jsi$ production mechanism and the $\jsi$ polarization by using the longitudinal momentum distribution of $\jsi$ mesons inside a fully reconstructed jet. The corresponding observable where a specific hadron is identified inside a jet is generically called the jet fragmentation function~\cite{Procura:2009vm,Jain:2011xz}. Experimentally, the distributions of hadrons inside jets have been measured at the LHC for light hadrons~\cite{Aad:2011sc,Chatrchyan:2014ava}, heavy mesons~\cite{Aad:2011td}, and more recently, for $\jsi$ mesons~\cite{Aaij:2017fak}. In particular, measuring the distribution of $\jsi$ mesons inside jets is an exciting opportunity for the following reasons. In comparison with the inclusive $\jsi$ cross section, i.e., the $p_T$ spectrum, the distribution inside the jet probes the $\jsi$ fragmentation function at a more differential level. Therefore, it should be possible to reveal detailed information about the non-perturbative hadronization process, which in turn should lead to new insights about the $\jsi$ production mechanism. The idea to measure $\jsi$ mesons inside jets was first proposed in~\cite{Baumgart:2014upa} in the context of exclusive $n$-jet processes. See also~\cite{Bain:2016clc,Bain:2016rrv}. In this work, we perform the calculation of $\jsi$ mesons in jets where the observable is defined to be inclusive over the entire final state except for the observed jet~\cite{Kaufmann:2015hma,Kang:2016mcy,Kang:2016ehg,Dai:2016hzf}. These types of (semi-) inclusive observables are easily accessible by the experiments and a direct comparison between theory and data is possible. The framework used in this work was derived within Soft Collinear Effective Theory (SCET)~\cite{Bauer:2000ew,Bauer:2000yr,Bauer:2001ct,Bauer:2001yt,Beneke:2002ph} and allows for the resummation of single logarithms in the jet size parameter $R$. In addition, we propose in this Letter for the first time to measure the polarization of $\jsi$ mesons inside jets. From our numerical analysis presented below, we find that this observable has even more discriminative power than the unpolarized cross section.

{\it Definition and factorization.} The distribution of $\jsi$ mesons within a fully reconstructed jet in proton-proton collisions, $p+p\to ({\rm jet} \, \jsi)+X$, is described by the so-called jet fragmentation function, denoted as $F^{\jsi}(z_h, p_T)$ which is defined as
\bea
F^{\jsi}(z_h, p_T) = \left.\frac{d\sigma^{\jsi}}{dp_T d\eta  dz_h}\right/\frac{d\sigma}{dp_T d\eta},
\eea
where we suppressed the $\eta$-dependence in $F^{\jsi}(z_h, p_T)$. The numerator and denominator are the differential cross sections of jets with and without the reconstruction of the $\jsi$ in the jet, while $\eta$ and $p_T$ are the jet rapidity and transverse momentum, respectively. Furthermore, $z_h=p_{\jsi}^+/p_{\rm jet}^+$ denotes the momentum fraction of the jet carried by the $\jsi$. The plus momentum is defined for any four vector $v^\mu$ as $v^+ = v^0 + v^z$ in a frame where the ``$z$''-axis is along the jet direction. The factorized form of the differential cross section for $\jsi$ production within a jet is given by~\cite{Kaufmann:2015hma,Kang:2016ehg}
\bea
\frac{d\sigma^{\jsi}}{dp_T d\eta  dz_h} = \sum_{a,b,c} f_a\otimes f_b\otimes H_{ab}^c\otimes {\mathcal G}_c^{\jsi}.
\label{eq:fac}
\eea
Here $\otimes$ denote convolution products over the partonic momentum fractions, $\sum_{a,b,c}$ represents the sum over all relevant partonic channels, and we have suppressed the arguments of the various functions for simplicity. See~\cite{Kang:2016ehg} for more details. The $f_{a,b}$ represent the parton distribution functions and $H_{ab}^c$ are the hard functions~\cite{Jager:2002xm}. The ${\mathcal G}_c^{\jsi}(z, z_h, p^+_{\rm jet}R, \mu)$ are the semi-inclusive fragmenting jet functions (siFJFs), which describe the fragmentation of a $\jsi$ meson inside a jet with radius $R$. The jet is initiated by a parton $c$ and carries a momentum fraction  $z = p_{\rm jet}^+/p_{c}^+$ of the outgoing parton. Note that we consider a cross section that is inclusive about the everything else in the final state besides the identified jet and its substructure~\cite{Kaufmann:2015hma,Kang:2016mcy}.

The siFJFs follow time-like DGLAP evolution equations, the same as those for the usual fragmentation
functions which describe the transition of a final state parton into a specific observed hadron~\cite{Kang:2016ehg}. By evolving the siFJFs through the DGLAP equations from their characteristic scale to the hard scale $\mu\sim p_T$, one can perform $\ln R$ resummation for narrow jets.
At the same time, the siFJFs describe the distribution of hadrons inside the jet, and, thus, contain important information about the hadronization of $\jsi$. In particular, ${\mathcal G}_i^{\jsi}$ can be expanded in terms of $\jsi$ fragmentation functions (FFs) as follows:
\bea
{\mathcal G}_i^{\jsi}(z, z_h, p_{\rm jet}^+R,\mu) =& \sum_j \int_{z_h}^1\frac{dz_h'}{z_h'} {\mathcal J}_{ij}(z, z_h/z_h', p_{\rm jet}^+R, \mu) 
\nonumber\\
&\hspace{-15mm}\times D_{j}^{\jsi}(z_h', \mu) + {\mathcal O}(m_{\jsi}^2/(p_{\rm jet}^{+}R)^2).
\label{eq:GD}
\eea
The coefficients ${\mathcal J}_{ij}$ were derived in~\cite{Kang:2016ehg}, where it was also shown that the natural matching scale should be $\mu_{\mathcal G}\sim p_T R$. 
Within the NRQCD formalism, the $\jsi$ FFs can be further factorized at an initial scale $\mu_0\sim m_{\jsi}$  with the following form
\bea
D_{i\to \jsi}(z_h', \mu_0) = \sum_{n} \hat d_{i\to [Q\bar Q(n)]}(z_h', \mu_0) \langle {\mathcal O}_{[Q\bar Q(n)]}^{\jsi}\rangle,
\label{eq:ffs}
\eea
where the summation runs over all intermediate non-relativistic $Q\bar Q$ states, labeled as $n =\,^{2S+1}L_J^{[1,8]}$, with superscript [1] (or [8]) denoting color singlet (or octet) state. The functions $\hat d_{i\to [Q\bar Q(n)]}$ are the short-distance coefficients and are perturbatively calculable within NRQCD, and have been derived in the past, see, e.g., Refs.~\cite{Ma:2013yla,Bodwin:2003wh}. On the other hand, $\langle {\mathcal O}_{[Q\bar Q(n)]}^{\jsi}\rangle$ are the non-perturbative NRQCD LDMEs. We use the calculated $\jsi$ FFs at an initial scale $\mu=3m_c$, and evolve them to the scale $\mu_{\mathcal G} = p_T R$ to be used in Eq.~\eqref{eq:GD}. 

{\it $\jsi$ polarization in the jet.} Besides measuring the $\jsi$ distribution in the jet, one can study the polarization of the produced $\jsi$. The polarization can be determined analogously to single inclusive $\jsi$ production, e.g., by measuring the angular distribution of the decay lepton pair $\ell^+\ell^-$ in the so-called helicity frame~\cite{Braaten:2014ata},
\bea
\frac{d\sigma^{\jsi(\to \ell^+\ell^-)}}{d\cos\theta}\propto 1+\lambda_F\cos^2\theta.
\eea
Here, $\lambda_F$ denotes the $J/\psi$ polarization measured in a jet, and $\lambda_F=1(-1)$ corresponds to a purely transversely (longitudinally) polarized $J/\psi$. Based on the factorization formalism in Eq.~\eqref{eq:fac}, $\lambda_F$ can be computed as follows:
\bea\label{eq:lambdaF}
\lambda_F(z_h, p_T)  = \frac{F^{\jsi}_T - F^{\jsi}_L}{F^{\jsi}_T + F^{\jsi}_L},
\eea
where $F^{\jsi}_{T,L}$ are the jet fragmentation functions for producing a $\jsi$ with transverse (or longitudinal) polarization. The total unpolarized jet fragmentation function is given by: $F^{\jsi} = 2F^{\jsi}_T + F^{\jsi}_L$. Since the $\jsi$ polarization is taken into account by the corresponding fragmentation functions, the $F^{\jsi}_{T,L}$ can be calculated using the same factorization formalism in Eq.~\eqref{eq:fac}. One only has to replace the unpolarized FFs $D_{i\to \jsi}$ in Eq.~\eqref{eq:ffs} by the polarized ones $D_{i\to \jsi}^{T, L}$. Note that the polarized FFs $D_{i\to \jsi}^{T, L}$ can be calculated within NRQCD analogously, 
\bea
D_{i\to \jsi}^{T,L}(z_h', \mu_0) = \sum_{n} \hat d_{i\to [Q\bar Q(n)]}^{T,L}(z_h', \mu_0) \langle {\mathcal O}_{[Q\bar Q(n)]}^{\jsi}\rangle.
\label{eq:D-pol}
\eea
The polarized short-distance coefficients for the states $^3S_1^{[1]}$, $^1S_0^{[8]}$, $^3S_1^{[8]}$, $^3P_J^{[8]}$ up to order of $\alpha_s^2$ are given in~\cite{Ma:2015yka} (For the state $^1S_0^{[8]}$, order of $\alpha_s^3$ contribution has been computed recently~\cite{Braaten} and Heavy quark to $^3S_1^{[1]}$ FF to next-to-leading order is computed in~\cite{Sepahvand:2017gup}, while the polarized short-distance coefficients for $g\to\,^3S_1^{[1]}$ were calculated in~\cite{kang} which first appear at order $\alpha_s^3$. 

{\it Phenomenology at the LHC.} We now present calculations for the $\jsi$ production and polarization within a fully reconstructed jet in proton-proton collisions at the LHC. We choose a center-of-mass energy of $\sqrt{s} = 7$ TeV, and assume that the jets are reconstructed through the anti-${\rm k_T}$ algorithm with a jet radius of $R=0.6$. For $\jsi$ production, we include all the relevant states: $^3S_1^{[1]}$, $^1S_0^{[8]}$, $^3S_1^{[8]}$, $^3P_J^{[8]}$. Thus, the results will depend on four NRQCD LDMEs: $\langle {\mathcal O}^{\jsi}(^3S_1^{[1]})\rangle$, $\langle {\mathcal O}^{\jsi}(^1S_0^{[8]})\rangle$, $\langle {\mathcal O}^{\jsi}(^3S_1^{[8]})\rangle$, and $\langle {\mathcal O}^{\jsi}(^3P_0^{[8]})\rangle$. These LDMEs have been fitted to $\jsi$ $p_T$-spectra by different groups which obtained very different values. Specifically we adopt the results from four groups: Bodwin {\it et.al.} in~\cite{Bodwin:2014gia}, Butenschoen {\it et.al.} in~\cite{Butenschoen:2011yh}, Chao {\it et.al.} in~\cite{Chao:2012iv}, and Gong {\it et.al.} in~\cite{Gong:2012ug}. See Table.~\ref{table:nrqcd} for the relevant numerical values. Below, the different fits will be referred to as Bodwin, Butenschoen, Chao, and Gong. 
 \begin{footnotesize}
\begin{table}[hbt]
\caption{$\jsi$ NRQCD LDMEs from four different groups.}
\label{table:nrqcd}
\begin{center}
  \begin{tabular}{  l |c | c | c | c }
  \hline
    & $\langle {\mathcal O}(^3S_1^{[1]})\rangle$ & $\langle {\mathcal O}(^1S_0^{[8]})\rangle$ & $\langle {\mathcal O}(^3S_1^{[8]})\rangle$ & $\langle {\mathcal O}(^3P_0^{[8]})\rangle$ \\ 
    & GeV$^3$ & $10^{-2}$ GeV$^3$ & $10^{-2}$ GeV$^3$ & $10^{-2}$ GeV$^5$ 
        \\ \hline
    Bodwin & 0~\footnote{Note: in \cite{Bodwin:2014gia}, the contribution from the $^3S_1^{[1]}$ state is assumed to be small and excluded from the fit.} & 9.9 & 1.1  & 1.1 \\ \hline
       Butenschoen & 1.32 & 3.04 & 0.16 & $-0.91$\\ \hline
   Chao & 1.16 & 8.9 & 0.30  & 1.26 \\ \hline
   Gong & 1.16 & 9.7 & $-0.46$ & $-2.14$ \\ \hline
  \end{tabular}
\end{center}
\end{table}
  \end{footnotesize}

In Fig.~\ref{fig:overall}, we plot the jet fragmentation function $F^{\jsi}(z_h, p_T)$ as a function of $z_h$ for three different jet transverse momentum $p_T$ bins: $[50,100]$, $[100, 150]$, $[150, 200]$ GeV. One finds that the LDMEs from different groups lead to very different results for $F^{\jsi}(z_h, p_T)$. For example, the parametrizations of Butenschoen and Gong can lead to a difference of almost an order of magnitude for the jet fragmentation function in the small $z_h$ region. This is caused mainly by the difference in signs of the LDMEs.
\bef
\includegraphics[width=2.8in]{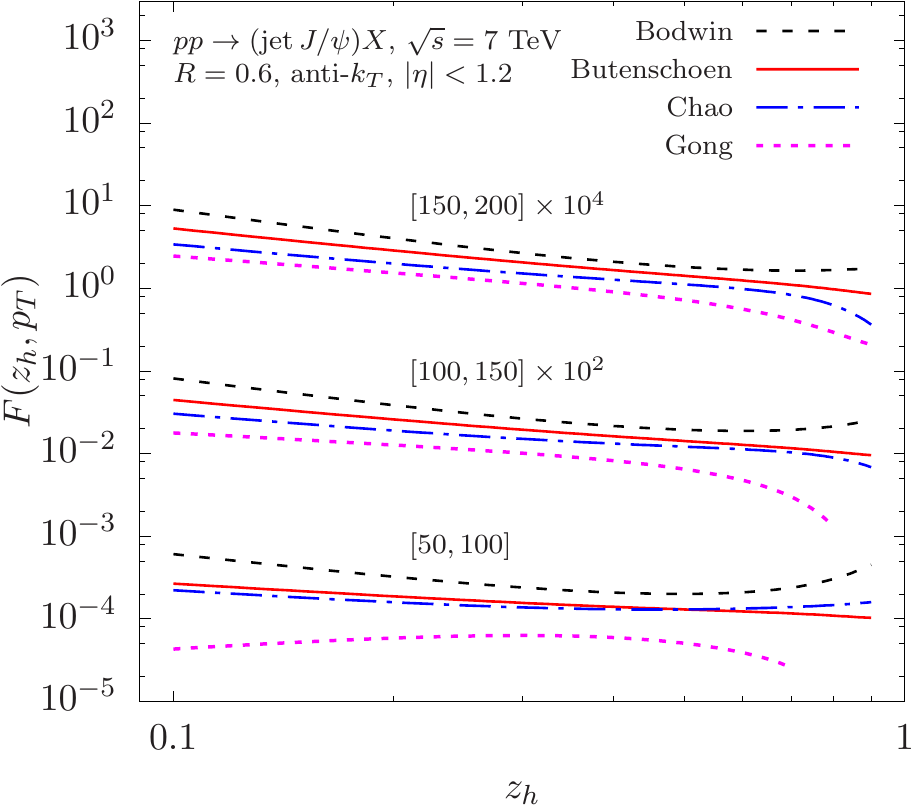}
\caption{The jet fragmentation function $F^{\jsi}(z_h, p_T)$ as a function of $z_h$ at $\sqrt{s}= 7$ TeV. Jets are reconstructed using the anti-${\rm k_T}$ algorithm with $R = 0.6$ and $|\eta| < 1.2$. The numbers in the square brackets correspond to different jet transverse momentum $p_T$ bins.}
\label{fig:overall}
\eef
The drastic difference between the $\jsi$-jet fragmentation function 
in Fig.~\ref{fig:overall}, evaluated with the LDMEs from different groups, precisely 
demonstrates that the $\jsi$ inclusive $p_T$-spectrum, as an inclusive observable, 
does not have the discriminative power 
to fully constrain the four relevant NRQCD LDMEs. However, as a more differential observable (in $z_h$), the $\jsi$-jet fragmentation function is a much more sensitive probe of these NRQCD LDMEs. The fact that the experimental measurements on the $\jsi$-jet fragmentation function at the LHC have already begun~\cite{Aaij:2017fak} is very encouraging. 

\bef
\includegraphics[width=2.8in]{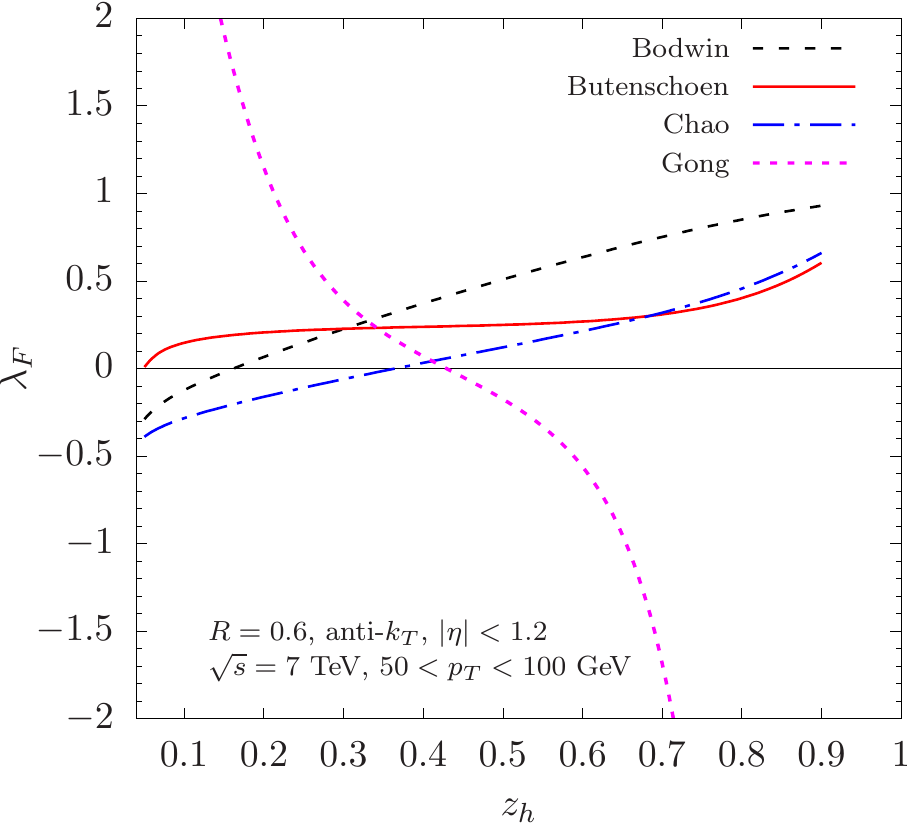}
\caption{The polarization of $\jsi$ mesons in the jet ($\lambda_F$) plotted as a function of $z_h$. The jet $p_T$ is integrated from 50 to 100 GeV.}
\label{fig:pol1}
\eef
To further explore the discriminative power of the $\jsi$ distribution in the jet, we study the polarization of $\jsi$ mesons in the jet, i.e., the observable $\lambda_F$ as defined in Eq.~(\ref{eq:lambdaF}). In Fig.~\ref{fig:pol1}, we show the result for $\lambda_F$ as a function of $z_h$, where the jet $p_T$ is integrated over the interval of $[50, 100]$ GeV. Again the parametrizations of different groups lead to distinctive predictions for the $\jsi$ polarization in the jet. For example, the Gong parametrization gives a transverse polarization $\lambda_F>0$ at small $z_h \lesssim 0.4$, which then becomes a longitudinal polarization $\lambda_F<0$ at large $z_h$. On the other hand, all three other parametrizations lead to a transverse polarization $\lambda_F > 0$ at large $z_h$, while the polarizations differ at small $z_h$ with very different magnitudes from that of the Gong parametrization. 

This vast difference shows once again the great discriminative power of $\jsi$-jet fragmentation functions, which is extremely good in terms of verifying NRQCD factorization formalism and constraining NRQCD LDMEs. 
It is instructive to emphasize that all the LDMEs were obtained so far by fitting only the data from the inclusive $J/\psi$ cross sections, and some of the fits need major cancellations between the production channels 
with different LDMEs. It is, then, entirely possible that the $J/\psi$ fragmentation functions which are expressed 
in terms of the same LDMEs, but with a very different combination of 
perturbative coefficients, can be negative or unphysical. The major cancellation 
obtained when fitting the $p_T$ distribution may not be satisfied when 
evaluating the $J/\psi$ fragmentation functions. This explains why we find 
$|\lambda_F|> 1$ when using the LDMEs of one particular fit at certain values 
of $z_h$ in Fig.~\ref{fig:pol1}.
In this sense, the $\jsi$-jet fragmentation functions can clearly lead to more stringent constraints on the LDMEs. In fact, one can even combine the usual $\jsi$ $p_T$-spectrum data with the $\jsi$-jet fragmentation function data to perform a joint global fit to extract NRQCD LDMEs. Such a global fit is expected to give much better constrained LDMEs, which would lead to more accurate information on heavy quarkonium formation.

To end this part, we discuss how our theoretical calculations with the 
existing LDMEs have resulted in the $\lambda_F$ as shown in Fig.~\ref{fig:pol1}.
The polarization $\lambda_F$ of a physical $\jsi$ is determined by the relative size of LDMEs, as well as the polarization properties for producing the four corresponding
{\it partonic} $[Q\bar Q(n)]$ states, which are determined by the perturbative coefficients $\hat d_{i\to [Q\bar Q(n)]}^{T,L}$ in Eq.~\eqref{eq:D-pol}. For these four relevant partonic  $[Q\bar Q(n)]$ states, we have: (1) $^1S_0^{[8]}$ with $J=0$ has no polarization preference. (2) the $^3S_1^{[1]}$ channel has a small preference toward a transverse polarization from our calculation. (3) $^3S_1^{[8]}$ has a strong preference toward a transverse polarization in the large $z_h$ region due to the contribution $\sim\delta(1-z_h)$ from the $g\to c\bar c$ fragmentation process. However, it leads to a longitudinal polarization in the small $z_h$ region due to DGLAP evolution. (4) The $^3P_J^{[8]}$ contribution tends to have a longitudinal polarization. 

With the knowledge of the polarization properties for producing the four {\it partonic} $[Q\bar Q(n)]$ states, given above, and 
the numerical values of the LDMEs summarized in Table.~\ref{table:nrqcd}, we are able to achieve a qualitative understanding of $\lambda_F$ for 
the production of 
the {\it hadronic} $\jsi$ state 
in Fig.~\ref{fig:pol1}. Taking the Butenschoen LDMEs as an example, in Fig.~\ref{fig:pol2}, we plot the individual 
contributions to $\lambda_F$ from different channels: $^3S_1^{[1]}$, $^1S_0^{[8]}$, $^3S_1^{[8]}$, $^3P_J^{[8]}$. Since the parametrization by Butenschoen has a positive value for the $\langle {\mathcal O}(^3S_1^{[8]})\rangle$ LDME, we obtain a transverse polarization ($\lambda_F > 0$) for almost all values of $z_h$, while it turns into a longitudinal polarization ($\lambda_F < 0$) at small $z_h$, consistent with the polarization properties of producing the partonic $Q\bar Q(^3S_1^{[8]})$ state.
On the other hand, a negative $\langle {\mathcal O}(^3P_0^{[8]})\rangle$, leads to a transverse polarization for the physical $\jsi$ production ($\lambda_F > 0$), which is opposite to  
the polarization contribution of producing
the partonic $Q\bar Q(^3P_J^{[8]})$ state. Therefore, one observes the additive and 
competing effects between the $^3S_1^{[8]}$ and $^3P_J^{[8]}$ contributions at large and
small $z_h$, respectively. The results for the other parametrizations in Fig.~\ref{fig:pol1} can be understood in a similar way.
\bef
\includegraphics[width=2.8in]{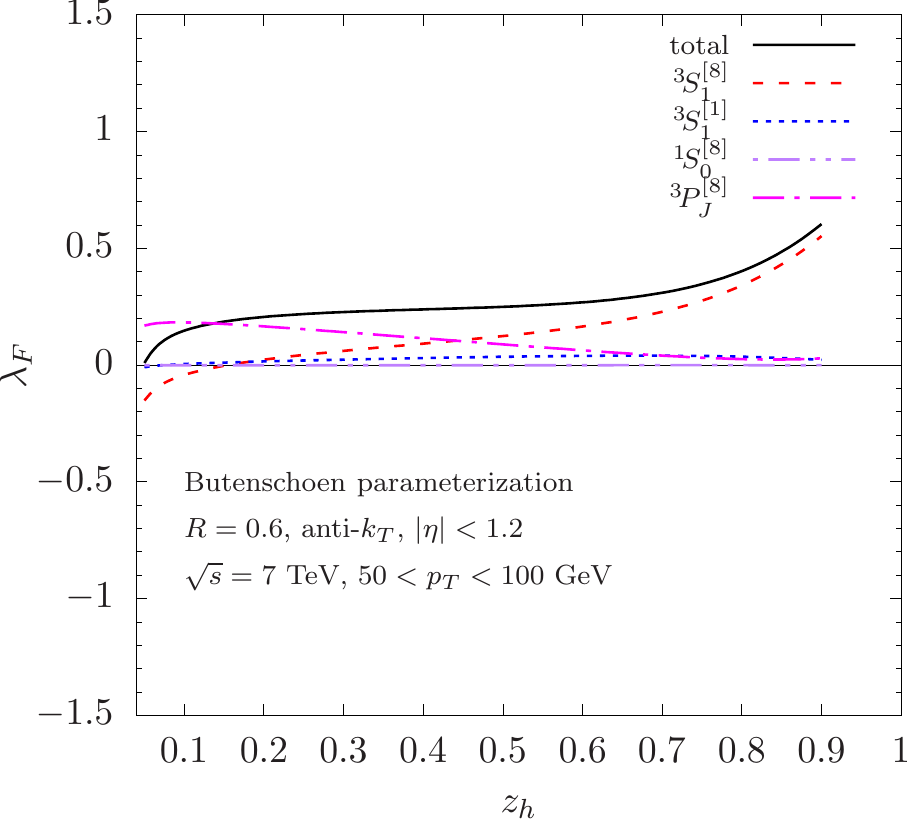}
\caption{The contributions to $\lambda_F$ from different channels: $^3S_1^{[1]}$, $^1S_0^{[8]}$, $^3S_1^{[8]}$, $^3P_J^{[8]}$ are plotted as a function of $z_h$. The NRQCD LDMEs are taken from Butenschoen {\it et.al.}~\cite{Butenschoen:2011yh}.}
\label{fig:pol2}
\eef 

This detailed analysis shows that the behavior and size of $\lambda_F$ in different $z_h$ regions is very sensitive to the short-distance coefficients as well as the different 
values of the LDMEs. Future measurements of the production and polarization of $\jsi$ mesons inside the jet will be very valuable to constrain the NRQCD formalism and LDMEs, and in turn provide unique information on the heavy quarkonium production mechanism. 

{\it Conclusions.} In this Letter, we studied the distribution and polarization of $\jsi$ mesons within a jet in proton-proton collisions at the LHC. Using a recently developed factorized formalism within SCET, we performed the resummation of single logarithms of the jet radius parameter $R$ up to next-to-leading logarithmic accuracy for the $\jsi$ distribution inside jets at LHC energies. We found that the NRQCD long-distance matrix elements extracted from a global analysis by four different groups lead to very different predictions for the $\jsi$ distribution inside the jet. Even though the parametrization of these four groups all describe the inclusive $\jsi$ cross section, the predicted $\jsi$ distribution inside the jet can differ by an order of magnitude for certain $z_h$ regions. We further defined an observable $\lambda_F$ which gives the polarization of $\jsi$ mesons in the jet. We found that this observable leads to even more discriminative power and thus, it can provide
better constraints for the LDMEs in global fits and more accurate information on the non-perturbative formation of heavy quarkonia. A complimentary study in~\cite{Bain:2017wvk} provided similar conclusions. 
We encourage the experimentalists to perform such measurements at the LHC and RHIC. We expect that these new observables will shed new light on the $\jsi$ production mechanism, and could likely lead to the eventual resolution of the $\jsi$ polarization puzzle. 

We thank Eric Braaten, Philip Ilten, Peter Jacobs, Rongrong Ma, Yan-Qing Ma, Thomas Mehen, and Michael Williams for helpful discussions. This work is supported by the National Science Foundation under Contract No. PHY-1720486 (Z.K.), the U.S. Department of Energy under Contract No.~DE-AC05-06OR23177 (J.Q.), No.~DE-AC02-05CH11231 (F.R.), No.~DE-FG02-91ER40684 (H.X.), No.~DE-AC02-06CH11357 (H.X.), and No.~DE-SC0011726 (H.Z.).

\bibliographystyle{h-physrev}
\bibliography{bibliography}

\end{document}